\documentclass[preprint]{elsarticle}

\makeatletter
\def\ps@pprintTitle{%
 \let\@oddhead\@empty
 \let\@evenhead\@empty
 \def\@oddfoot{}%
 \let\@evenfoot\@oddfoot}
\makeatother

\usepackage{amsmath,amssymb,amsfonts}
\usepackage{graphicx}
\usepackage{url}
\usepackage{hyperref}
\usepackage{breakurl}

\usepackage{lineno,hyperref}
\modulolinenumbers[5]

\date{\empty}

\begin{document}

\begin{frontmatter}

\title{A Relationship Between SIR Model \\ and Generalized Logistic Distribution \\ with Applications to SARS and COVID-19}

\author{Hideo Hirose}
\address{Bioinformatics Center, Kurume University \\   
Kurume, Fukuoka, Japan}
\ead{hirose\_hideo@kurume-u.ac.jp}

\begin{abstract}
This paper shows that the generalized logistic distribution model is derived from the well-known compartment model, consisting of susceptible, infected and recovered compartments, abbreviated as the SIR model, under certain conditions.
In the SIR model, there are uncertainties in predicting the final values for the number of infected population and the infectious parameter. 
However, by utilizing the information obtained from the generalized logistic distribution model, we can perform the SIR numerical computation more stably and more accurately. Applications to severe acute respiratory syndrome (SARS) and Coronavirus disease 2019 (COVID-19) using this combined method are also introduced.
\end{abstract}


\end{frontmatter}

\section{Introduction}

The well-known compartment model, consisting of susceptible, infected and recovered compartments, abbreviated as the SIR model, has been commonly used in infectious disease spread simulations for more than half a century although the mathematical model is very simple  (see Kermack and McKendrick, 1933; Anderson and May, 1991; Diekmann and Heesterbeek, 2000, e.g., for general descriptions). Such a long life-length proves the effectiveness of this model, resulting in a variety of expansions and many actual applications (see Berge et al., 2017; Chowell et al., 2006; Ferguson et al., 2001; Vaidya et al., 2014, e.g., for specific applications).
Similar to other cases, this model has been recently applied to Coronavirus disease 2019 (COVID-19) caused by the novel severe acute respiratory syndrome coronavirus 2 (SARS-CoV-2); see Wu et al., 2020; Tuite et al., 2020; Dehning et. al., 2020; Fang et al., 2020; Fokas et al., 2020; Giordano et al., 2020; Anand et al., 2020; Bertozzi et al., 2020. 

Meanwhile, the generalized logistic distribution has also been used in statistical infectious disease spread predictions (see e.g., Zhou and Yan, 2003; Hirose, 2007) as an empirical model. The method is also used in COVID-19 case (see Aviv-Sharon and Aharoni, 2020, e.g.).

However, 
there seems to be no plain explanations connecting this differential equation model and the probability distribution model.
In this paper, firstly,
we show that the generalized logistic distribution model can be derived from the SIR model under certain conditions.
Then, we can propose a more accurate and stable prediction method by combining these two models.
Applications to severe acute respiratory syndrome (SARS) and COVID-19 using this combination method are also introduced.

\section{SIR Model and Its Uncertainties in Numerical Computations}

\subsection{SIR Model with Cumulative Infected Population}

The SIR model uses the ordinary differential equations 
\begin{eqnarray}
\displaystyle  {dS(t) \over dt}&=&-\lambda S(t) I(t), \label{eq:SIR1} \\ 
\displaystyle  {dI(t) \over dt}&=&\lambda S(t) I(t) - \gamma  I(t),  \label{eq:SIR2}  \\
\displaystyle  {dR(t) \over dt}&=&\gamma  I(t), \label{eq:SIR3}
\end{eqnarray}
where $S$, $I$, and $R$ mean the susceptible, infectious, and removed populations, and the parameters $\lambda$, and $\gamma$ are the infection rate, and removal rate (recovery rate). 
In the SIR model, for example, a person could change his or her condition from susceptible to infected with a ratio $\lambda$, then to removed with a ratio $\gamma$. Removed persons will never become susceptible. From Equations (\ref{eq:SIR1}), (\ref{eq:SIR2}), (\ref{eq:SIR3}), we have
\begin{eqnarray}
\displaystyle  {dS(t) \over dt} + {dI(t) \over dt} + {dR(t) \over dt} =0,
 \label{eq:SIR4} 
 \end{eqnarray}
which means $S(t)+I(t)+R(t)=const.$ This is a total population size, and we denote this by $N$.
Here, we introduce $T(t)=I(t)+R(t)$ for further discussions. This is the number of cumulative infected persons.

\subsection{Importance of Reproduction Number}

Assuming that $I(t)$ is small enough comparing to $S(t)$ at early stages, that is, $S(t) \approx S(0) \approx N$, then Equation (\ref{eq:SIR2}) can be written
\begin{eqnarray}
\displaystyle  {dI(t) \over dt}&=&\lambda S(0) I(t) - \gamma  I(t) = (\lambda S(0) - \gamma) I(t).  \label{eq:SIR5} 
\end{eqnarray}
This fundamental ordinary differential equation can be easily solved as 
\begin{eqnarray}
\displaystyle  I(t)&=& I(0) e^{ (\lambda S(0) - \gamma) t }, \label{eq:SIR6} 
\end{eqnarray}
using integration, where $I(0)$ and $S(0)$ are the initial number of infected persons and the initial number of susceptible persons, respectively.
If we define 
$R_0={\lambda S(0) / \gamma}  \approx {\lambda N / \gamma} $ 
which is called the basic reproduction number, $I(t)$ becomes increasing when $\displaystyle R_0 > 1$, and decreasing when $\displaystyle R_0 < 1$. 
That is, $R_0$ plays an important role in determining the pandemic phenomena or extinction phenomena at early stages of the infectious disease spread.
Therefore, to know $R_0$ as early as possible is considered to be crucial as well as to find the solutions for the SIR model.
By rewriting $R_0={\lambda S(0) / \gamma}$ to $R_0=({\lambda S(0) I(t)) / (\gamma} I(t))$, it is worth mentioning that the condition that the number of inflow of infected persons is equivalent to the number of outflow of removed persons results in $R_0=1$.  Later, this concept is useful to consider the current reproduction number $R_c(t)$ at current time $t$ such that
\begin{eqnarray}
\displaystyle  R_c(t)&=&{\lambda (t) S(t) \over \gamma (t)} 
,  \label{eq:R_c} 
\end{eqnarray}
where $\lambda (t)$ and $\gamma (t)$ are infection rate and removal rate at time $t$, respectively.

\subsection{Uncertainties of SIR Numerical Computations}

In solving the SIR ordinary differential equations as an initial value problem, we require parameter values of $\lambda$ and $\gamma$, and initial values of $S(0)$, $I(0)$ and $R(0)$. 

\subsubsection{Solving Difference Equations}

By observation, we can obtain daily populations for $I(t)$ and $R(t)$ because daily numbers of newly infected, died and recovered (cured) persons are noticed in public. We note that $I(t)$ is not identical to the daily number of newly infected persons.
Then, parameters $\lambda$ and $\gamma$ for the SIR model at time $t$ can be roughly obtained by using the simultaneous difference equations below, regarding the differential equations as the difference equations.
\begin{eqnarray}
\displaystyle  \lambda(t)&=&{S(t) - S(t+1) \over S(t) I(t)},  \label{eq:SIR7} \\
\displaystyle  \gamma(t)&=&{R(t+1) - R(t) \over I(t)} \label{eq:SIR8}.  
\end{eqnarray}

Since daily values of $\lambda(t)$ and $\gamma(t)$ are unstable to some extent, such an instability shall be removed by adopting the mean values computed from the latest $\lambda(t)$ and $\gamma(t)$ values; for example, seven days average values can be used. However, this is not an optimal solution.
To find much more accurate estimates for parameters $\lambda$ and $\gamma$, we may use the best-backward solution (BBS) method explained below. 

\subsubsection{Best-bakward solution, BBS}

First, we obtain the initial guesses of $\lambda^{(0)}$ and $\gamma^{(0)}$ for $\lambda$ and $\gamma$, e.g., by using the simultaneous difference equations above. 
Then, we estimate the optimum values of $\hat {\lambda}$ and $\hat {\gamma}$ for $\lambda$ and $\gamma$ by using the simplex method (Nelder and Mead, 1965), which is shown to be an extension from the method previously proposed (Hirose, 2012). In optimization, we evaluate the following function $E(n,s)^{(k)}$ iteratively,
\begin{eqnarray}
E(n,s)^{(k)} &=& \sum_{j={n-s}}^n  \{ ({T}^{(k)}(t_j) - \tilde{T}(t_j))^2 + ({R}^{(k)}(t_j) - \tilde{R}(t_j))^2 \}, \label{eq:BBS1} \\
                   &=& \sum_{j={n-s}}^n  \{ ({S}^{(k)}(t_j) - \tilde{S}(t_j))^2 + ({R}^{(k)}(t_j) - \tilde{R}(t_j))^2 \}, \label{eq:BBS2} 
\end{eqnarray}
where $\tilde{T}(t_j)$, $\tilde{S}(t_j)$ and $\tilde{R}(t_j)$ are the numbers of observed values for cumulative infected persons, susceptible persons and removed persons; ${T}^{(k)}(t_j)$ and ${R}^{(k)}(t_j)$ are $k$-th iterative solutions for the numbers of cumulative infected persons, susceptible persons and removed persons of the ordinary differential equations of the SIR. We continue this iteration until $|E(n,s)^{(k+1)} - E(n,s)^{(k)}| < \varepsilon$ holds, where $\varepsilon$ is a small positive number. In solving the SIR equations, the solutions are obtained backward from time $t=t_n$ to time $t=t_{n-s}$ somehow, e.g., Runge-Kutta method. Finally, we could find the converged values $\hat {\lambda}$ and $\hat {\gamma}$. We note that ${T}^{(k)}(t_n) = \tilde{T}(t_n)$, ${S}^{(k)}(t_n) = \tilde{S}(t_n)$ and ${R}^{(k)}(t_n) = \tilde{R}(t_n)$ for all $k$. Taking into account the importance of the most recent observed values, we often use $s=7$.

However, $S(t)$ becomes unreliable because it strongly depends on $N$, and we cannot determine an appropriate size of $N$.
It may be a small district size, or a large country size, depending on the region of disease spread.
That is, $N$ is unknown in general because plausible uninfected persons who can contact infected persons are not identified.
It is an uncertainty factor in using the SIR model that we cannot determine $N$.
There may be a resolution to such an inconvenience by removing the term $N$ from Equations (\ref{eq:SIR1}), (\ref{eq:SIR2}) and (\ref{eq:SIR3}). That is, we assume $S(t)+I(t)+R(t)=1$. However, we often want to know the actual infected population size, but not the ratio of infected population to the total population $N$. We will deal with such the case.

\subsubsection{Uncertainty Factors in the SIR}

To look at this phenomenon, we have performed a simulation study using the SARS case in Hong Kong in 2003 (SARS case data, (2003)). We assume cases that $N$ is 2,000 (strongly restricted area population), 10,000, 100,000, 1,000,000, and 6,810,000 (actual Honk Kong population in 2003). 

\begin{figure}[htbp]
\begin{center}
\includegraphics[height=6cm]{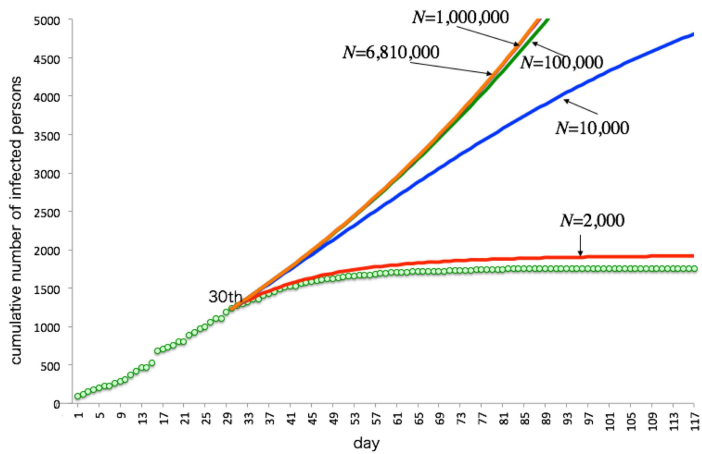}
\end{center}
\caption{Various cases of the predicted curves for the number of cumulative infected persons, $T(t)$, after 30th day using from 22nd to 30th day observed data via the SIR model (SARS in Honk Kong in 2003). We assumed cases that $N$ is 2,000, 10,000, 100,000, 1,000,000, and 6,810,000.}
\label{fig:SARSSIR30thT}
\end{figure}

Figure \ref{fig:SARSSIR30thT} shows the predicted curves for the number of cumulative infected persons, $T(t)$, after 30th day using observed data from 22nd to 30th day via the SIR model; here, day 1 was March 17, 2003. In the figure, the observed values corresponding to $T$ are the  superimposed using dotted points. We see that the prediction curves strongly depend on $N$. The larger the value of $N$, the steeper the increasing tangent. In the cases of $N \ge 100,000$, the curves show blow-up at the moment of 30th day, although $T(t)$ is bounded above by the upper limit $N$. Therefore, we can mention that we cannot predict the robust value for $T$ if $N$ is unknown, typically at early stages.

There is also another uncertainty factor in solving the SIR ordinary differential equations.
We cannot estimate the consistent value for ${\lambda} (t)$ unless $N$ is clearly predetermined, typically at early stages.

\section{The Generalized Logistic Distribution and the SIR}

The generalized logistic distribution (GLD) model developed by Richards (see Richards, 1959) can be applied to flexible growth function for empirical use. This  model is based on a simpler model (see Von Bertalanffy, 1938) to describe the increase of weight as a function of the metabolism process of animals.
The GLD is applied also to other fields such as hydrology (see Zakaria et al., 2012, e.g.), medical fields
in infectious disease spread modelling such as SARS, foot and mouth disease (FMD), Zika virus disease, Ebola virus disease, and SARS Cov-2 and in growth modelling such as physiochemical phenomenon, psychological issues, survival time of diagnosed leukemia patients, and weight gain data.

\subsection{Generalized Logistic Distribution}

The three-prameter generalized logistic distribution function is defined as
\begin{eqnarray}
\displaystyle 
F(t; \sigma, \mu, \beta)
={1 \over \{ 1+\exp(-{t-\mu \over \sigma}) \} ^{\beta} }, 
\label{eq:GLD1} 
\end{eqnarray}
where $\sigma$, $\mu$ and $\beta$ denote the scale, location, and shape parameters, respectively.
By introducing $z={(t - \mu) / \sigma}$,
we have the standard generalized logistic distribution expressed by
\begin{eqnarray}
\displaystyle 
F(z; \beta)
={1 \over \{ 1+\exp(-z) \}^{\beta} },
\label{eq:GLD2} 
\end{eqnarray}
where
\begin{eqnarray}
\displaystyle 
\exp(-z)
= \exp(-{t - \mu \over \sigma})
= \exp({\mu \over \sigma}) \exp(-{t \over \sigma}). 
\label{eq:GLD3} 
\end{eqnarray}

In estimating the parameters $\theta=(\sigma, \mu, \beta)^T$, we often use the maximum likelihood estimation method. Since observed data are usually daily data, the likelihood function $L(\theta)$ can be constructed by using the grouped truncated model expressed as
\begin{equation}
L(\theta) 
= \left\{ { F(t_{0};\theta)  \over F(t_{n};\theta) } \right\} ^ {k_0}
\cdot 
\prod_{i=1}^{n}
 \left\{ { F(t_{i};\theta) - F(t_{i-1};\theta) \}  \over F(t_{n};\theta) } \right\} ^ {k_i},
\end{equation}
where $k_i \ (0 \le i \le n)$ represents the number of infected persons from time $-\infty$ to $t_{0}$ or $t_{i-1}$ to $t_{i}$.
When the total number of cases is known in advance, we can also use the trunsored model (Hirose, 2007).

\subsection{Derivation of GLD from the SIR}

In the SIR model (\ref{eq:SIR1}), (\ref{eq:SIR2}), (\ref{eq:SIR3}), we have assumed that $I(t)$ is small enough comparing to $S(t)$ at early stage, i.e., $S(0) \approx N$, then we have derived the simple ordinary differential equation (\ref{eq:SIR4}) with the solution of (\ref{eq:SIR5}). This solution shows the explosive increasing population for the infected persons when $R_0>1$.
In the real world, the number of infected persons is bounded above. Thus, a much more realistic model is required at later stages in disease spreading.

Since $T(t)=N-S(t)$, from Equation (\ref{eq:SIR1}), we have
\begin{eqnarray}
\displaystyle {dT(t) \over dt} =  - {dS(t) \over dt} = \lambda S(t) I(t) = \lambda S(t) (T(t) - R(t)).
\label{eq:GLD4} 
\end{eqnarray}
Assuming that $R(t)=0$, i.e., none of infected persons will be transferred to the removed population, then, this equation becomes
\begin{eqnarray}
\displaystyle {dT(t) \over dt} = \lambda T(t) (N - T(t)). 
\label{eq:GLD5} 
\end{eqnarray}
This equation shows a symmetry between $T(t)$ and $(N - T(t))$, and consequently, the solution represents increasing flat S-shaped curve.
By integration, we can easily derive the solution of (\ref{eq:GLD5}) as 
\begin{eqnarray}
\displaystyle T(t) = { N \over 1 +  ({ N  \over T(0)} - 1) e^{- \lambda N t}},
\label{eq:GLD6} 
\end{eqnarray}
which is called the logistic function.
The inflection point becomes $(\log(N/T(0)-1) / \lambda N, N/2)$ by solving 
${d^2 T(t) / dt^2} = 0$.

In the real world, this model is still unrealistic because asymmetric curves are often observed. 
Then, we assume that $({T(t) / N})$ shall be $({T(t) / N})^m$ because it would be natural to think that the susceptible persons would be more/less affected by infected persons depending on the magnitude of $m$ rather than linearly affected. For example, if half of the population is already infected, the ratio of infectious persons will not be half but will be inflated to $3/4$ when $m=2$, and it will be shrunk to $(2 -  \sqrt{2}) / 2$ when $m=1/2$.

Thus, we assume the following ordinary differential equation
\begin{eqnarray}
\displaystyle {dT(t) \over dt} = b T(t) (1 - ({T(t) \over N})^m),
\label{eq:GLD7} 
\end{eqnarray}
where $b$ is a constant holding that $b = \lambda N $ when $m=1$.
To solve this equation, firstly, we set $y(t)={T(t) / N}$, and further, we use change of variables such that $z(t)=(y(t))^{-m}$ (refer to Skiadas, 2009, e.g.,  for such transformations). Then, Equation (\ref{eq:GLD7}) can be written as 
\begin{eqnarray}
\displaystyle {dz \over dt} = - b m (z - 1),
\label{eq:GLD8} 
\end{eqnarray}
which can also be solved by integration, and 
the solution is
\begin{eqnarray}
T(t) = {N \over  \{1 + (({T(0) \over N})^{-m} - 1) \exp (-bmt) \} ^{1/m} }. 
\label{eq:GLD7}
\end{eqnarray}

This reveals that the curve of $T(t)$ shows the same shape to one expressed by Equation (\ref{eq:GLD2}) except for the scale.
Therefore, the generalized logistic distribution is derived from the SIR model with certain assumptions.
These assumptions could be applied to the cases at early stages of the disease spread, we may use this probability distribution as the statistical model representing the infectious disease spread phenomena.
The time $t$ for the inflection point becomes $(\log((({N / T(0)})^{m} - 1)/m)) / bm$ by solving 
${d^2 T(t) / dt^2} = 0$.

We have the relationships of parameters between Equation (\ref{eq:GLD1}) and Equation (\ref{eq:GLD7}) such that 
$b = {\beta / \sigma}$, 
$k = (1 + \exp({\mu / \sigma}))^{-{\beta} }$, 
$m = {1 / \beta}$, 
$\sigma  =  {1 / (bm)}$,
$\mu  = ({\log (k^{-m}-1)) / (bm)} $, 
$\beta  = {1 / m}$.

\subsection{Estimation of $N$}

\begin{figure}[htbp]
\begin{center}
\includegraphics[height=6cm]{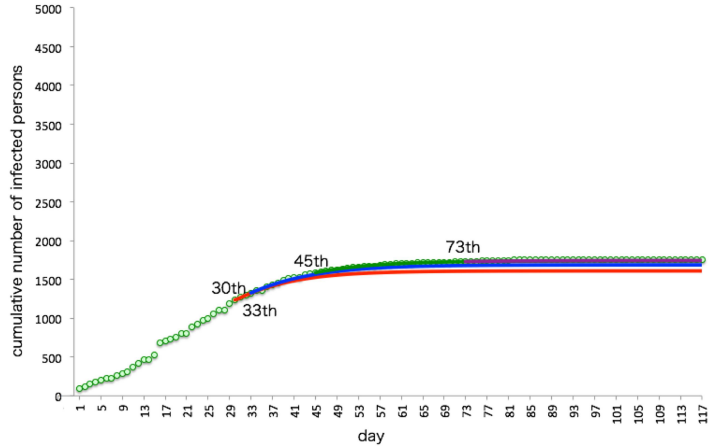}
\end{center}
\caption{Predicted curves for the number of cumulative infected persons, $T(t)$, after 30th, 33rd, 45th and 73th day using observed data from the first day to the last observed day via GLD model (SARS in Honk Kong in 2003).}
\label{fig:SARSGLT} 
\end{figure}

Using exactly the same example in the previous section, we show, in Figure \ref{fig:SARSGLT}, the predicted curves for the number of cumulative infected persons, $T(t)$, after 30th, 33rd, 45th and 73th day using observed data from the first day to the last observed day, by maximizing the likelihood function (\ref{eq:GLD14}). Dotted points show the observed values for $T$. In this GLD model, we can estimate the final (i.e., $t \rightarrow \infty$) value for $T(t)$ such that
\begin{eqnarray}
\hat{T}(\infty) =  {T(t) \over F(t;\hat{\theta})}, \label{eq:GLD14}
\end{eqnarray}
where $F(t;\hat{\theta})$ expresses the cumulative distribution function value using estimated parameter $\hat{\theta}$ at time $t$.
Looking at the figure, we see that the prediction curves for $T(t)$ are close to the observed values, although the prediction curves show underestimated results to some extent. 

In addition, the GLD model can predict the final value $\hat{T}(\infty) = N$ even at early stages.
It will not make sense that we compare the final value $\hat{T}(\infty)$ using the GLD model with that using the SIR model because $\hat{T}(\infty)$ using the SIR model does not provide consistent values. Therefore, we compare the value $\hat{T}(t_{\rm conv})$ using the GLD model with that using the SIR model, where $t_{\rm conv}$ expresses the time when $\hat{T}(t)$ seems to converge. In the SARS case example, we set $t_{\rm conv} = 117$.
We introduce the two terms of $\hat{T}(t_{\rm conv})_{\rm GLD}(t)$ and $\hat{T}(t_{\rm conv})_{\rm SIR}(t)$; the former represents predicted $\hat{T}(t_{\rm conv})$ estimated at truncation time $t$ using the GLD model, and the latter $\hat{T}(t_{\rm conv})$ at time $t$ using the SIR model. 

\begin{figure}[htbp]
\begin{center}
\includegraphics[height=6cm]{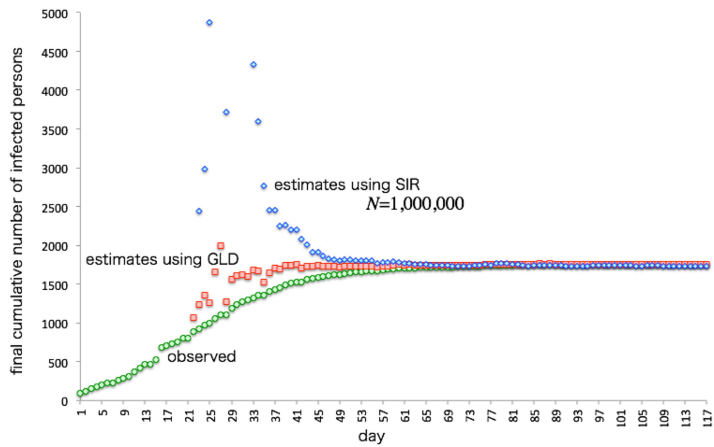}
\end{center}
\caption{Comparison of the L-plots between $\hat{T}(117)_{\rm GLD}(t)$ and $\hat{T}(117)_{\rm SIR}(t)$ (SARS in Honk Kong in 2003).}
\label{fig:L-plot} 
\end{figure}

We define L-plot such that time $t$ locates in horizontal axis, and that $\hat{T}(t_{\rm conv})$ locates in the vertical axis. 
Figure \ref{fig:L-plot} shows the comparison between $\hat{T}(t_{\rm conv})_{\rm GLD}(t)$ and $\hat{T}(t_{\rm conv})_{\rm SIR}(t)$ at various time $t$; in the figure, observed $T(t)$ are superimposed. 
Although we have selected a rather small value of $117$ for $t_{\rm conv}$, the estimated value of $\hat{T}(117)_{\rm SIR}(t)$ shows unstable behavior than $\hat{T}(117)_{\rm GLD}(t)$ does. 
Therefore, we consider using the information for $N$ using the GLD model, and to combine the GLD model use and the SIR model use next.

\section{Combination Method}

As described above, we can predict the number of cumulative infected persons after the last day of observation using the SIR model under appropriate conditions. However, to keep the estimates reliable, we should pay attention to provide adequate parameter values, in particular for infection parameter $\lambda$ and total population $N$. Otherwise, even if we can roughly obtain the current reproduction number $R_c$, we cannot know consistent estimates for $\lambda$ and $N$. 

On the contrary, we do not require the information of $N$ to estimate the parameter $\theta$ in the GLD model. In addition, we can estimate $N$ even at early stages. Thus, in order to obtain the more accurate estimates of the parameters in the SIR model, we may utilize that information from the GLD model. This is called the combination method. Using these two models simultaneously, we can expect to make more accurate predictions. 

\subsection{Application to COVID-19 case}

\begin{figure}[htbp]
\begin{center}
\includegraphics[height=8.5cm]{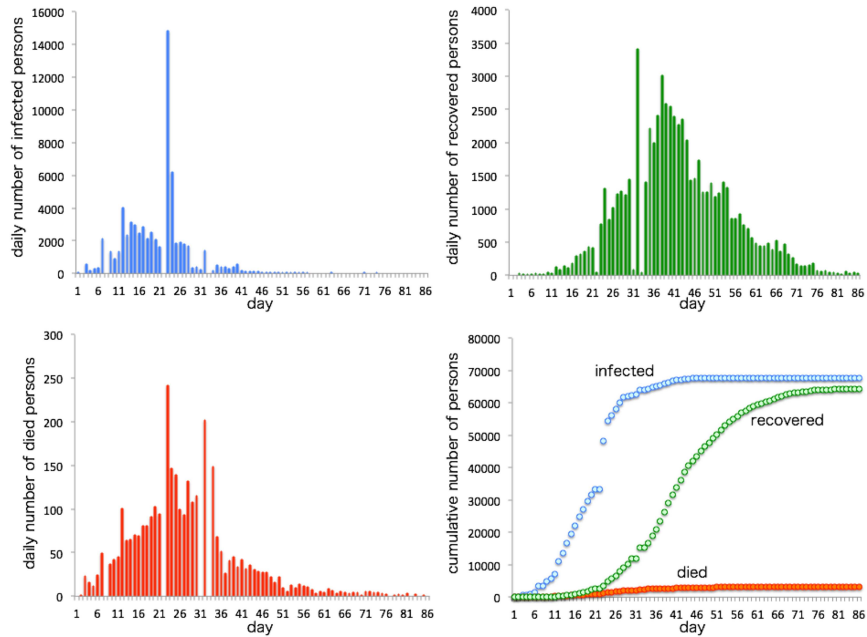}
\end{center}
\caption{Observed COVID-19 case data in Hubei in 2020. Daily number of infected persons, daily number of died persons, daily number of removed persons, and cumulative numbers of them are illustrated}
\label{fig:HubeiData} 
\end{figure}

Figure \ref{fig:HubeiData} shows the observed COVID-19 case data in Hubei in 2020 (Hubei case data, (2020)). Looking at the figure, we see that the infection spreads very quickly, but recovered slowly. From the infection to recover, it took three weeks referring the median time of infection curve and recovered curve. 

\begin{figure}[htbp]
\begin{center}
\includegraphics[height=5cm]{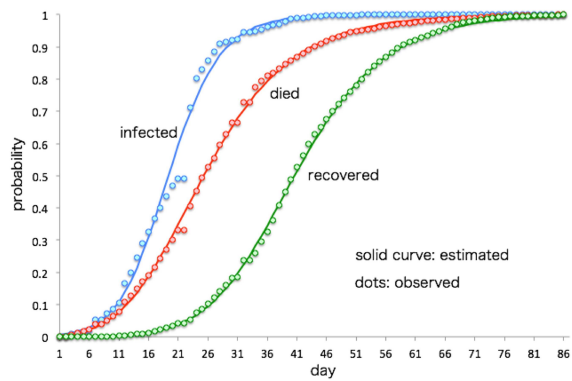}
\end{center}
\caption{Estimated cumulative distribution functions, Cdfs, for infected, died and removed cases, fitted to COVID-19 case data in Hubei in 2020. Solid curves show estimated Cdfs, and dots show the observed values.}
\label{fig:HubeiProbability} 
\end{figure}

First, we have fitted the generalized logistic distributions to observed values of cumulative number of infected persons, cumulative number of died persons and cumulative number of recovered persons. Figure \ref{fig:HubeiProbability} shows the cumulative distribution functions and corresponding observed data, i.e., the cumulative number of persons are divided by the total number of persons. We can see that the three observed cases are well fitted to the generalized logistic distributions. According to Hirose, 2007, among the generalized lognormal, generalized extreme-value, generalized gamma and generalized logistic distributions, the generalized logistic distribution showed the best-fit model. 

\begin{figure}[htbp]
\begin{center}
\includegraphics[height=6cm]{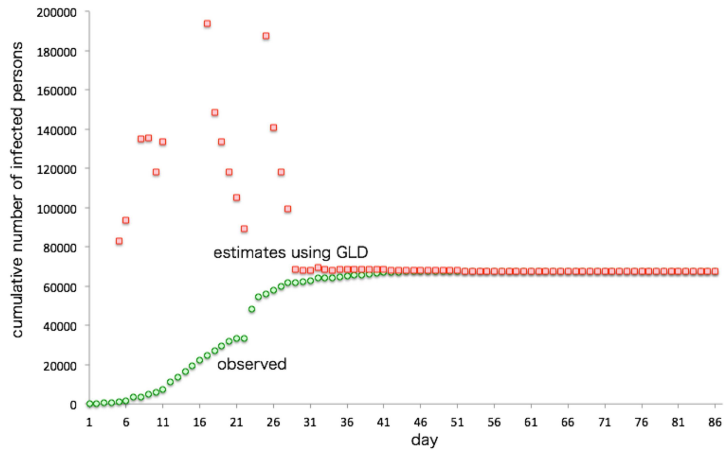}
\end{center}
\caption{L-plot for COVID-19 case data in Hubei in 2020. Observed values are superimposed using dots.}
\label{fig:HubeiLplot} 
\end{figure}

To the observed number of infected persons, we have applied the L-plot in Figure \ref{fig:HubeiLplot}, assuming that the observed values follow generalized logistic distributions. From the figure, we see that the estimates for $N$ seem to stable after 29th day from the beginning, and it seems to converge around $70,000$. Thus, we set $N$ to $70,000$ in the SIR model.

\begin{figure}[htbp]
\begin{center}
\includegraphics[height=5.5cm]{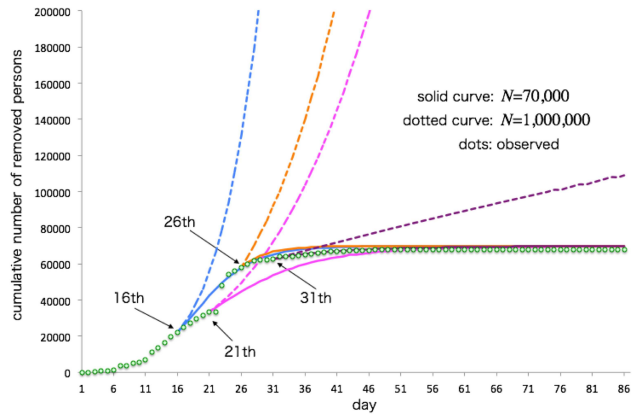}
\end{center}
\caption{Predicted curves of various cases for the number of cumulative infected persons after the last day of observation for COVID-19 case data in Hubei in 2020. Observed values are superimposed using dots.}
\label{fig:HubeiPrediction} 
\end{figure}

Figure \ref{fig:HubeiPrediction} shows the predicted curves after the last day of observation. In the figure, solid curves express the case of $N=70,000$, and dotted curves $N=1,000,000$ for the sake of comparison. Clearly, the case of $N=1,000,000$ is misleading because there are large discrepancies between the predicted values and observed values. On the contrary, it seems that the predicted values are close to observed values in the case of $N=70,000$. 

\section{Concluding Remarks}

Although the SIR model has been actively used for a long time and has been useful for prediction, there are uncertainties in predicting the final values for the number of infected population and the infectious parameter. In this paper, we have introduced that the generalized logistic distribution model can be derived from the SIR model under certain conditions. In using the generalized logistic distribution model, we can resolve one of the uncertainty factors, resulting in the use of such the information for the numerical computations in the SIR model. We have proposed a more accurate and stable prediction methodology by cooperating these two models with each other. Applications to SARS and COVID-19 using this combined method are also introduced.


\section*{References}

\begin{thebibliography}{10}
\expandafter\ifx\csname url\endcsname\relax
  \def\url#1{\texttt{#1}}\fi
\expandafter\ifx\csname urlprefix\endcsname\relax\def\urlprefix{URL }\fi
\expandafter\ifx\csname href\endcsname\relax
  \def\href#1#2{#2} \def\path#1{#1}\fi

\baselineskip 10pt 

\bibitem{AndersonMay}
Anderson, R. M., May, R., 1991. Infectious diseases of humans: Dynamics and control,
  Oxford University Press.

\bibitem{Anand}
Anand, N., Sabarinath, A., Geetha S., Somanath, S., 2020. Predicting the spread of Covid‐19 using SIR model augmented
  to incorporate quarantine and testing, Transactions of the Indian National
  Academy of Engineering 5, 141--148.

\bibitem{Berge}
Berge, T., Lubuma, J.M.-S., Moremedi, G.M., Morris, N., Kondera-Shava, R., 2017. A simple mathematical model for ebola in africa, Journal of Biological Dynamics, 11(1), 42--74.

\bibitem{Bertalanffy}
Von Bertalanffy, L., 1938. A quantitative theory of organic growth (inquiries on
  growth laws. ii), Human Biology, 10(2), 181--213.

\bibitem{Bertozzia}
Bertozzi, A. L., Franco, E., Mohler G., Short, M. B., Sledge D., 2020. The challenges of modeling and forecasting the spread of
  covid-19, PNAS. 117(29), 16732--16738.

\bibitem{Chowell}
Chowell, G., Rivas, A., Hengartner, N., Hyman,, Castillo-Chavez, C., 2006. Critical response to post-outbreak vaccination against
  foot-and-mouth disease, AMS Contemporary Mathematics, 47, 73--87.

\bibitem{Dehning}
Dehning, J., Zierenberg1, J., Spitzner, F. P., Wibral, M.,Neto, J. P., Wilczek, M., Priesemann V., 2020. Inferring change points in the spread of covid-19 reveals
  the effectiveness of interventions, Science, 369~(eabb9789), 1--9.

\bibitem{DiekmannHeesterbeek}
Diekmann, O., Heesterbeek, J. A. P., 2000. Mathematical epidemiology of infectious diseases:
  model building, analysis and interpretation, New York: Wiley.

\bibitem{Fang}
Fang, Y., Nie, Y.,  Penny, M., 2020. Transmission dynamics of the covid-19 outbreak and
  effectiveness of government interventions: A data-driven analysis, J Med
  Virol., 645--659.

\bibitem{Ferguson}
Ferguson, N., Donnelly, C. A., Anderson, R. M.   (2001. The foot-and-mouth epidemic in Great Britain: Pattern of
  spread and impact of interventions, Stochastic Analysis and Applications, 292
 1155--1160.

\bibitem{Fokas}
Fokas, A. S., Dikaios, N., Kastis, G. A., 2020. Mathematical models and deep learning for predicting the
  number of individuals reported to be infected with Sars-Cov-2, J. R. Soc.
  Interface, 17 1--12.

\bibitem{Giordano}
Giordano, G., Blanchini, F., Bruno, R., Colaneri, P., Di Filippo, A., Di Matteo, A., Colaneri, M., 2020. Modelling the Covid-19 epidemic and implementation of
  population-wide interventions in Italy, Nature Medicine, 26(2)
  855--860.

\bibitem{Hirose2007}
Hirose, H., 2007. The mixed trunsored model with
  applications to Sars, Mathematics and Computers in Simulation, 74(6)
  443--453.
  
\bibitem{Hirose2012}
Hirose, H., 2012. Estimation of the number of failures in the Weibull model using the ordinary differential equation, European Journal of Operational Research, 223(3), 722--731.
  
\bibitem{HubeiData}
Hubei case data, 2020.
  \url{https://github.com/CSSEGISandData/COVID-19/tree/master/csse_covid_19_data/csse_covid_19_daily_reports}.

\bibitem{KermackMcKendrick}
Kermack, W. O., McKendrick, A. G., 1933. Contributions to the mathematical theory of
  epidemics-iii. further studies of the problem of endemicity, in: Proceedings
  of the Royal Society, 94--122.

\bibitem{NelderMead}
Nelder, J. A., Mead, R., 1965. A simplex method for function minimization, The Computer
  Journal, 7, 308--313.

\bibitem{Richards}
Richards, F., 1959. A flexible growth function for empirical use, Journal of
  Experimental Botany, 10(29), 290--300.

\bibitem{SARSData}
SARS case data, 2003. \url{http://www.who.int/csr/sars/country/en/}.

\bibitem{SharonAharoni}
Aviv-Sharon, E., Aharoni, A., 2020. Generalized logistic growth modeling of the Covid-19
  pandemic in Asia, Infectious Disease Modelling, 5, 502--509.

\bibitem{Skiadas2010}
Skiadas, C., 2009. Exact solutions of stochastic differential equations: Gompertz,
  generalized logistic and revised exponential, Methodology and Computing in
  Applied Probability, 12(2), 261--270.

\bibitem{Tuite}
Tuite, A. R., Fisman, D. N., Greer, A. L., 2020. Mathematical modelling of Covid-19 transmission and
  mitigation strategies in the population of ontario, Canada, CMAJ, 192(19),
 E497--505.

\bibitem{Vaidya}
Vaidya, N. K., Morgan, M., Jones, T., Miller, L., Lapin, S.\& Schwartz, E. J., 2014. Modelling the epidemic spread of an h1n1 influenza outbreak
  in a rural university town, Epidemiology and Infection, 143(8),
  1610--1620.

\bibitem{Wu}
Wu, K., Darcet, D., Wang Q., Sornette, D., 2020. Generalized logistic growth modeling of the Covid-19 outbreak:
  comparing the dynamics in the 29 provinces in china and in the rest of the
  world, Nonlinear Dynamics, 1--21.

\bibitem{Zakaria}
Zakaria, Z. A., Shabri, A., Ahmad, U. N., 2012. Estimation of the generalized logistic distribution of
  extreme events using partial l-moments, Hydrological Sciences Journal, 57(3),
 424--432.

\bibitem{ZhouYan}
Zhou, G., Yan, G., 2003. Severe acute respiratory syndrome epidemic in Asia,
  Mathematics and Computers in Simulation, 9(12), 1608--1610.


\end{thebibliography}
\end{document}